\documentclass{emulateapj}

\newcommand{\oviii}{\ion{O}{8} $\lambda 18.97$}
\newcommand{\sixiv}{\ion{Si}{14} $\lambda 6.18$}

\newcommand{\fexviia}{\ion{Fe}{17} $\lambda 16.78$}
\newcommand{\mgxii}{\ion{Mg}{12} $\lambda 8.42$}

\shorttitle{X-rays line profiles in EX Hydrae.}
\shortauthors{Luna et al.}

\begin{document}


\title{Photoionized features in the X-ray spectrum of EX Hydrae}


\author{G. J. M. Luna\altaffilmark{1}, J. C. Raymond\altaffilmark{1}, N. S. Brickhouse\altaffilmark{1}, C. W. Mauche\altaffilmark{2}, D. Proga\altaffilmark{3}, D. Steeghs\altaffilmark{4}, R. Hoogerwerf\altaffilmark{5}}
\altaffiltext{1}{Harvard-Smithsonian Center for Astrophysics, 60 Garden St., Cambridge, MA, 02138, USA. E-mail: \texttt{gluna@cfa.harvard.edu}; \texttt{jraymond@cfa.harvard.edu}; \texttt{nbrickhouse@cfa.harvard.edu}}
\altaffiltext{2}{Lawrence Livermore National Laboratory, L-473, 7000 East Avenue, Livermore, CA 94550, USA. E-mail: \texttt{mauche@cygnus.llnl.gov}}
\altaffiltext{3}{Department of Physics and Astronomy, University of Nevada, 4505 South Maryland Parkway, Box 454002, Las Vegas, NV, 89154-4002, USA. E-mail: \texttt{dproga@physics.unlv.edu}}%
\altaffiltext{4}{Department of Physics, University of Warwick, Coventry CV4 7AL, UK. E-mail: \texttt{dsteeghs@cfa.harvard.edu}}
\altaffiltext{5}{Interactive Supercomputing, Inc., 135 Beaver Street, Waltham, MA 02452, USA. E-mail: \texttt{hoogerw@pobox.com}}




\begin{abstract}
We present the first results from a long (496 ks) {\it Chandra\/} High Energy Transmission Grating observation of the intermediate polar EX Hydrae.  
In addition to the narrow emission lines from the cooling post-shock gas, for the first time we have detected a broad component in some of the X-ray emission lines, namely \oviii , \mgxii , \sixiv , and \fexviia .
The broad and narrow components have widths of $\sim 1600~\rm km~s^{-1}$ and $\sim 150~\rm km~s^{-1}$, respectively. 
We propose a scenario where the broad component is formed in the pre-shock 
accretion flow, photoionized by radiation from the post-shock flow.
Because the photoionized region has to be close to the radiation source in order to produce strong photoionized emission lines from ions like \ion{O}{8}, \ion{Fe}{17}, \ion{Mg}{12}, and \ion{Si}{14}, our photoionization model constrains the height of the standing shock above the white dwarf surface. Thus, the X-ray spectrum from EX~Hya manifests features of both magnetic and non-magnetic cataclysmic variables. 
\end{abstract}




\section{Introduction }
\label{sec:intro}
EX Hydrae (EX Hya) belongs to a sub-class of magnetic cataclysmic variables (CVs), known as intermediate polars (IPs), wherein the white dwarf (WD) magnetic field ($\approx 0.1$--10 MG) channels the accreting material from the inner edge of the truncated accretion disk through the accretion curtains and accretion columns to spots near the magnetic poles \citep{rosen88}. In this model, the magnetically channeled material reaches highly supersonic velocities in the pre-shock flow ($v_{\rm ff}= [2GM_{\rm WD}/R_{\rm WD}]^{1/2} \simeq 6000~\rm km~s^{-1}$, where $v_{\rm ff}$ is the free-fall velocity for a WD of mass $M_{\rm WD}=0.79~M_{\odot}$ and radius $R_{\rm WD}=7.1\times 10^8$ cm) before passing through a strong stand-off shock near the WD surface. At the shock, the kinetic energy is converted into thermal energy, heating the gas to 
temperatures of $\sim 20$ keV 
\citep{fujimoto,masses}. Below the shock, in the post-shock flow, the cooling material radiates its thermal and residual gravitational energy through free-free, bound-bound, and free-bound radiation, which is primarily detected in X-rays.

The nature of the observed X-ray spectrum of CVs depends primarily on the specific accretion rate (i.e., the accretion rate per unit area). The small spot size and hence high specific accretion rate of magnetic CVs impose a conical geometry that does not allow the X-ray photons to escape without further interaction. Thus the observed X-ray spectrum is dominated by emission lines formed in a region photoionized by the radiation field from the shocked region. In contrast, the boundary layer at the inner edge of the accretion disk of non-magnetic CVs covers a relatively large area and therefore the specific accretion rate is low, allowing the X-ray photons to escape freely. \citet{koji} observed this ``binomial'' distribution in moderately exposed ($\sim 100$ ks) observations obtained with {\it Chandra\/} using the High Energy Transmission Grating (HETG), and generally found that the X-ray spectra of magnetic CVs are compatible with photoionized emission while those of non-magnetic CVs are consistent with
a collisionally ionized, cooling gas (known as a cooling-flow model in X-ray spectral fitting packages such as XSPEC\footnote{\url{http://heasarc.gsfc.nasa.gov/docs/xanadu/xspec/}}). Interestingly, EX Hya was the ``exception'' in this binomial distribution: its X-ray spectrum was well fit with a cooling-flow model. This can be explained if EX Hya has a tall shock \citep{allan98}, a lower accretion rate, and/or larger accretion spots than other IPs, and therefore a low specific accretion rate, leading to a dominant collisionally ionized, cooling spectrum.


We have obtained a deeper observation of EX Hya with the {\it Chandra\/} HETG in order to study a number of accretion-related phenomena. In this Letter, we present new results from this observation discussed in Section \ref{sec:obs}. In Section \ref{sec:profiles} we present an analysis of the X-ray line profiles that require a broad component, and a photoionization model to explain it. Discussion and conclusions are presented in Section \ref{sec:discussion}.

\section{Observations }
\label{sec:obs}

EX Hya was observed with {\it Chandra\/} using the HETG in combination with the ACIS-S for 496 ks. The observation was obtained in four segments (ObsIDs 7449: start time 2007 May 13 22:15:35 UT, exposure time 130.65 ks; 7452: start time 2007 May 17 03:12:38 UT, exposure time 49.17 ks, 7450: start time 2007 May 18 21:56:57 UT, exposure time 162.73 ks; and 7451: start time 2007 May 21 14:15:08 UT, exposure time 153.07 ks). We extracted High-Energy Grating (HEG) and Medium-Energy Grating (MEG) $\pm $ first-order spectra and Ancillary Responce Matrices (ARFs) using the CIAO\footnote{{\it Chandra\/} Interactive Analysis of Observations (CIAO v. 4.1)} script \texttt{fullgarf}, while Response Matrix Functions (RMFs) were extracted using \texttt{mkgrmf} script. We then combined ARFs and spectra from each observation segment using the \texttt{add\_grating\_spectra} script. For \sixiv\ and \mgxii , fits were performed using the HEG and MEG $\pm $ first order data, whereas for \fexviia\ and \oviii , fits were performed using only the MEG $\pm $ first order data, due to the rapid decay of the HEG effective area for $\lambda \gtrsim 17$~\AA\ ({\it Chandra\/} Proposer's Observatory Guide v11.0).

\section{Emission line profiles}
\label{sec:profiles}

\subsection{Line profiles analysis}
\label{subsec:profiles}

\begin{figure*}
\begin{center}
\includegraphics[scale=1.0]{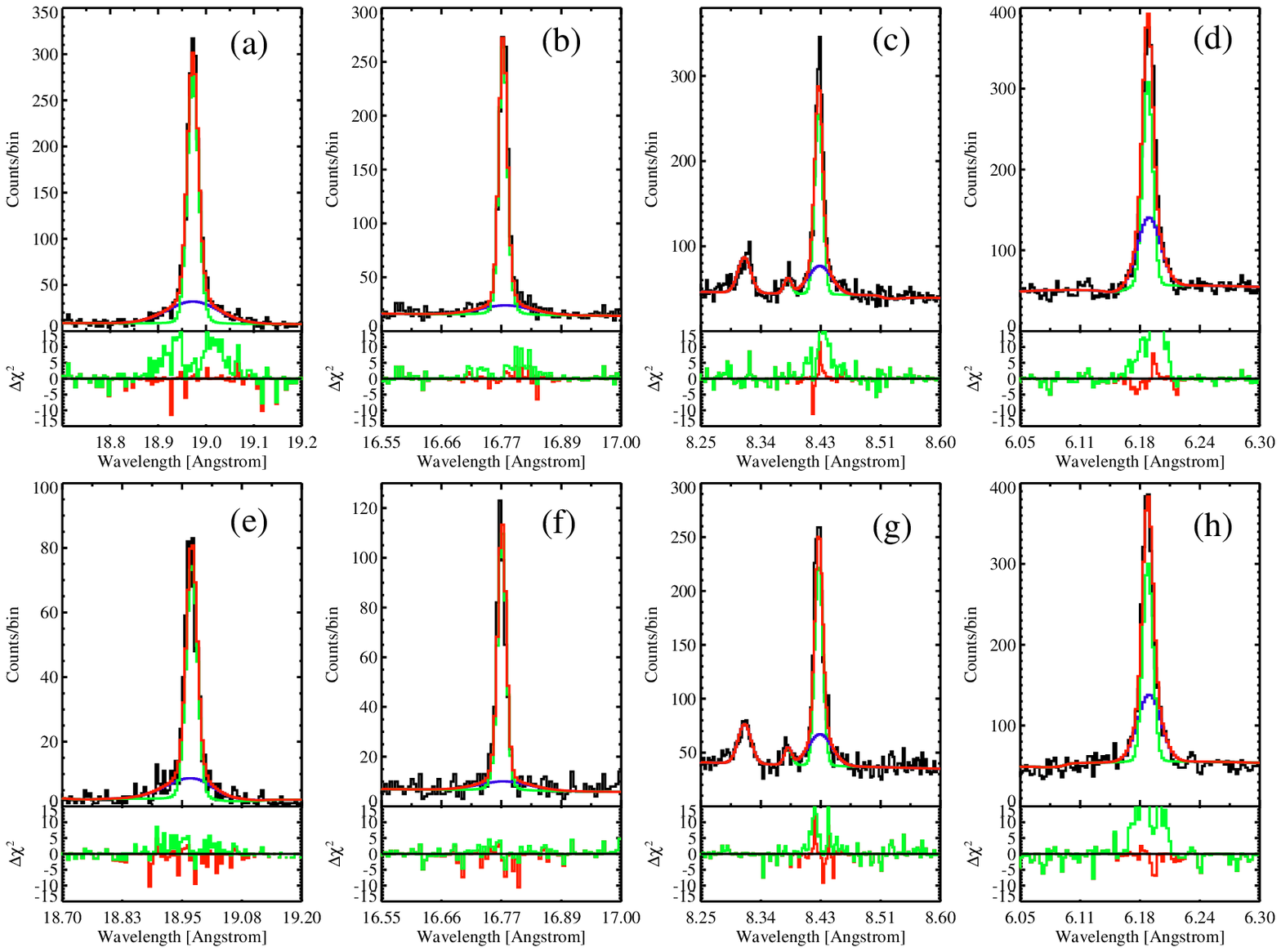}
\end{center}
\caption{Observed emission line profiles with overlaid double-Gaussian models of
({\it a\/}, {\it e\/}) \oviii\   (MEG $\pm 1$),
({\it b\/}, {\it f\/}) \fexviia\ (MEG $\pm 1$),
({\it c\/}, {\it g\/}) \mgxii\   (HEG $\pm 1$), and
({\it d\/}, {\it h\/}) \sixiv\   (HEG $\pm 1$). Data are shown by the black histograms and the narrow, broad, and net components of the model profiles are shown by the green, blue, and red histograms, respectively. The two emission lines (Fe XXI--XXIV $\lambda 8.31$ and \ion{Fe}{24} $\lambda 8.38$) in the vicinity of \mgxii\ were included in the fit. Below each panel we plot the residuals relative to the net double-Gaussian model (red) and to its narrow component (green). $\Delta\chi^2$ stands for the difference between the data and the model, squared, divided by the variance, with the sign of the difference between the data and the model.}
\label{fig:2}
\end{figure*}

In order to test models of the cooling of the post-shock gas \citep[e.g.][]{aizu,canalle}, we measured the fluxes of the strong emission lines observed in the {\it Chandra\/} HETG spectrum using a Gaussian to represent each emission line and a first-order polynomial to represent the nearby continuum. We found that some of the strong emission lines were poorly fit with such a model, and that acceptable fits to these lines, in particular \oviii , \fexviia , \mgxii , and \sixiv , could be obtained by adding a second Gaussian to represent the broad line wings. All the fit parameters were free to vary and their limits determined to 90\% confidence. Figure \ref{fig:2} shows the studied emission lines together with the best-fit models. Central wavelength, full width half maximum (FWHM), and observed flux of the narrow and broad components of each line are listed in Table \ref{table:1}.

\begin{figure}
\includegraphics[width=7cm,height=6.5cm]{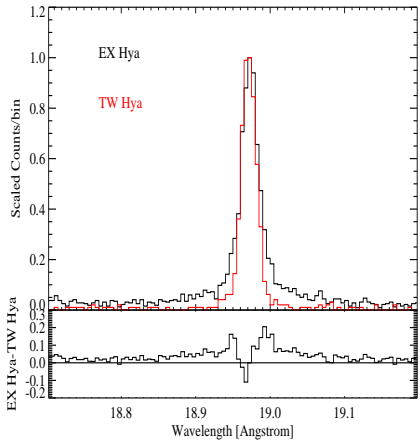}
\caption{{\it Upper panel:\/} Observed \oviii\ emission line profile in EX~Hya (black) and TW~Hya (red), scaled to the same peak intensity. {\it Lower panel:\/} Relative difference between the EX~Hya and TW~Hya profiles, showing the excess emission in the wings of the EX~Hya line profile. The lack of a broad component in the TW~Hya line profile demonstrates that the broad component seen in EX~Hya is not of instrumental origin.}
\label{fig:4}
\end{figure}

An instrumental origin for these broad emission lines seems unlikely. First, the broad component is detected in four spectral orders simultaneously (HEG and MEG $\pm $ first orders), whereas any broadening in the dispersion direction of a diffraction arm should not affect the dispersion in another arm. Second, as shown in Figure \ref{fig:4}, a comparison with a similarly deeply exposed observation of the pre-main-sequence star TW~Hya \citep[exposure time: 489 ks,][]{nancy} with the same instrument configuration does not show similar broad features.

\begin{figure*}
\includegraphics[scale=1]{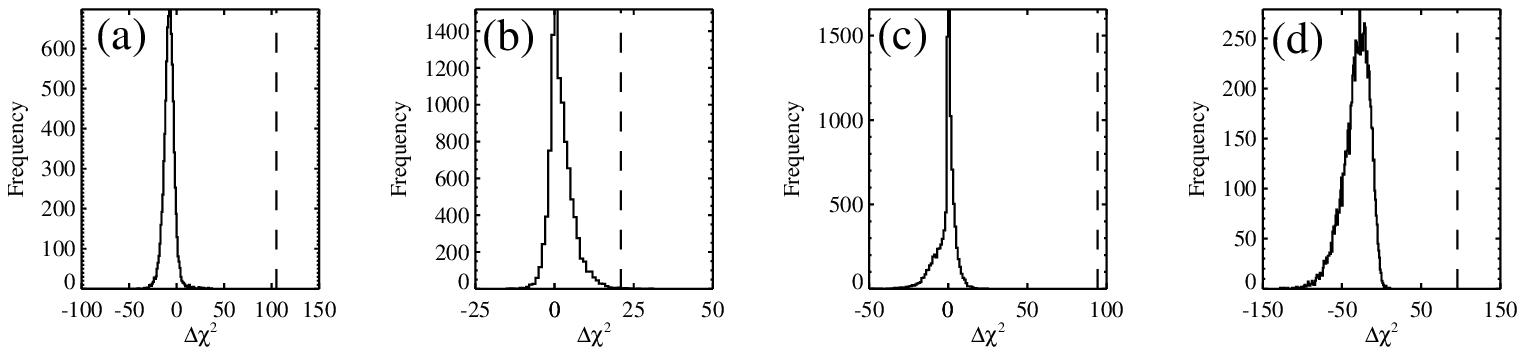}
\caption{Histograms of the distribution of $\Delta\chi^2$ for ({\it a\/}) \ion{O}{8}, ({\it b\/}) \ion{Fe}{17}, ({\it c\/}) \ion{Mg}{12}, and ({\it d\/}) \ion{Si}{14} resulting from fitting 10,000 Monte Carlo simulated spectra with a continuum plus (i) a single narrow Gaussian and (ii) narrow and broad Gaussians. In the case of \ion{O}{8}, \ion{Mg}{12}, and \ion{Si}{14}, only 1 out of 10,000 randomly generated spectra produce a $\Delta\chi^2$ (i.e. difference in $\chi^2$ between the fit of the double- and single-Gaussian models) equivalent to that found in our data; for \ion{Fe}{17}, 10 out 10,000 produce a $\Delta\chi^2$ equivalent to that found in our data. The $\Delta\chi^2$ value measured in the observed data is marked in each panel by a dashed line.}
\label{fig:ftest}
\end{figure*}

The statistical significance of the presence of a broad component in the line profiles of \oviii , \fexviia, \mgxii, and \sixiv\ in the HETG spectra of EX Hya was assessed by fitting 10,000 Monte Carlo realizations of the spectra in the neighborhood of each line, using a model consisting of a continuum plus a
single narrow Gaussian, with the same exposure time, response, and ancillary matrices as our observation, following the method described in \citet{protassov} and \citet{reeves}. We fit the simulated spectra using two models: a continuum plus (i) a single narrow Gaussian [with a total of 4 free parameters, resulting in: $\chi^{2}$(\ion{O}{8})=503.70; $\chi^2$(\ion{Mg}{12})=1671.69; $\chi^2$(\ion{Si}{14})=868.04, and $\chi^2$(\ion{Fe}{17})=162.72] and (ii) narrow and broad Gaussians [with a total of 7 free parameters, resulting in: $\chi^{2}$(\ion{O}{8})=398.60; $\chi^2$(\ion{Mg}{12})=1577.30; $\chi^2$(\ion{Si}{14})=772.40 and $\chi^2$(\ion{Fe}{17})=141.77]. The difference in $\chi^2$ between the fit of the double- and single-Gaussian models to the data was computed for each spectrum in order to test if the broad component could be produced purely by random, Poisson noise. We found that only 1--10 in 10,000 of the randomly generated single-Gaussian spectra could produce a similar decrease in $\chi^2$ when compared to the observed data, after the addition of a broad emission line (see Table \ref{table:1} and Figure \ref{fig:ftest}). The null hypothesis probability that the broad component could arise purely from Poisson noise is $\lesssim0.1$\%; thus the broad components are detected at 99.9\% confidence.

\begin{deluxetable*}{lccccccc}
\tablecolumns{8}
\tablewidth{0pc}
\tablecaption{Line Fit Parameters.}
\tablehead{
\colhead{} & \multicolumn{3}{c}{Narrow Component} & \colhead{} & \multicolumn{3}{c}{Broad Component} \\
             \cline{2-4}                                         \cline{6-8} \\
\colhead{} & \colhead{$\lambda $\tablenotemark{b}} & \colhead{FWHM\tablenotemark{c}} & \colhead{}& \colhead{} & \colhead{$\lambda $\tablenotemark{b}} & \colhead{FWHM\tablenotemark{c}} &  \colhead{}\\
\colhead{Line\tablenotemark{a}} & \colhead{(\AA )} & \colhead{$\rm (km~s^{-1})$}& \colhead{Flux\tablenotemark{d}} & \colhead{}& \colhead{(\AA )} & \colhead{$\rm (km~s^{-1})$}& \colhead{Flux\tablenotemark{d}} }
\startdata
\hbox to 1.0in{\oviii\leaders\hbox to 0.5em{\hss.\hss}\hfill}  & $18.972_{-0.006}^{+0.006}$ & $315_{-25}^{+34}$ & $7.13_{-0.3 }^{+0.4 }$  & &$18.972_{-0.011}^{+0.011}$ & $1922_{-162}^{+ 230}$ & $2.77_{-0.3 }^{+0.3 }$ \\
\hbox to 1.0in{\fexviia\leaders\hbox to 0.5em{\hss.\hss}\hfill}& $16.777_{-0.012}^{+0.024}$ & $117_{-50}^{+42}$ & $2.75_{-0.15}^{+0.18}$  & &$16.783_{-0.018}^{+0.019}$ & $1963_{-511}^{+1247}$ & $0.55_{-0.21}^{+0.19}$ \\
\hbox to 1.0in{\mgxii\leaders\hbox to 0.5em{\hss.\hss}\hfill}  & $ 8.422_{-0.006}^{+0.006}$ & $360_{-34}^{+41}$ & $1.08_{-0.08}^{+0.09}$  & &$ 8.424_{-0.006}^{+0.006}$ & $1329_{-108}^{+ 176}$ & $0.53_{-0.09}^{+0.10}$\\ 
\hbox to 1.0in{\sixiv\leaders\hbox to 0.5em{\hss.\hss}\hfill}  & $ 6.183_{-0.006}^{+0.006}$ & $410_{-57}^{+23}$ & $1.09_{-0.10}^{+0.03}$  & &$ 6.185_{-0.006}^{+0.006}$ & $1297_{- 78}^{+  69}$ & $0.84_{-0.05}^{+0.03}$ \\
\enddata
\tablenotetext{a}{\ion{O}{8} and \ion{Fe}{17} were fit using only the MEG arm; \ion{Mg}{12} and \ion{Si}{14} were fit using both the HEG and MEG arms.}
\tablenotetext{b}{Absolute wavelength accuracy: $\rm HEG=0.006$ \AA , $\rm MEG=0.011$ \AA . Relative wavelength accuracy: $\rm HEG=0.0010$ \AA , $\rm MEG=0.0020$ \AA\ ({\it Chandra\/} Proposer's Observatory Guide v11.0)}
\tablenotetext{c}{FWHM $\approx 2.3548 \sigma $.}
\tablenotetext{d}{Units of $10^{-4}$ photons cm$^{-2}$ s$^{-1}$.}
\label{table:1}
\end{deluxetable*}

\subsection{The photoionization model}
\label{sec:model}

The width of the broad component allows us to constrain the region where the line is formed. As listed in Table \ref{table:1}, we measured large widths in the broad components, which imply that the radiating plasma is moving with large velocities ($v\gtrsim 2000~\rm km~s^{-1}$). Of the three regions of the magnetically controlled accretion flow --- the accretion curtains, where the flow rises up and away from the truncated accretion disk; the pre-shock flow, where the relatively cool flow falls with high velocity toward the WD surface; and the post-shock flow, where the hot, low-velocity flow settles onto the WD surface --- the pre-shock flow is the most likely source of the observed broad components of the X-ray emission lines of EX Hya.


In principle either photoionization or collisional excitation could produce the observed broad components; however, if the width is interpreted as thermal broadening, the implied temperature is too high.
The FWHM of the emission line can be translated into temperature through:
\begin{equation}
T_{\rm ion}=\mu \times \left(\frac{\Delta \lambda_{\rm FWHM}}{7.7\times10^{-5} \times \lambda}\right)^2
\label{eq:thermal}
\end{equation}
where $T_{\rm ion}$ is the ion temperature in eV, $\mu$ is the ion mass, $\Delta \lambda_{\rm FWHM}$ is the observed FWHM of the line and $\lambda $ is the observed central wavelength. The measured widths imply temperatures $T_{\rm ion} \sim 80$ keV, much higher than the peak electron temperature measured of $\sim 20$ keV \citep{fujimoto,masses}. 

On the other hand, a photoionization origin for the observed broad emission lines is possible if there is a region in the system that fullfills certain conditions.
First, ionization parameters ($\xi=L_{\rm X}/nr^2$, where $L_{\rm X}$ is the X-ray luminosity, $n$ is the number density, and $r$ is the distance between the source and target of the photoionizing flux) of a few hundred are required to account for the presence of \ion{O}{8}, \ion{Mg}{12}, and \ion{Si}{14} \citep[e.g.,][]{kallman}. Second, the emission measure ($EM=n^2V$, where $V$ is the volume) of the photoionized plasma must be large enough to produce the observed flux.

The conditions above exclude the accretion curtains and favor the pre-shock photoionized flow as the source of the broad X-ray emission lines. Consider first the accretion curtains: using the observed \ion{O}{8} luminosity at an inner disk radius of $10\, R_{\rm WD}$ \citep[e.g.,][]{beuerman03} and with 
an ionization parameter of a few hundred results in a density of around $3\times 10^{10}~\rm cm^{-3}$, and hence a volume of $2\times 10^5\, R_{\rm WD}^3$, which is hundreds of times the volume available at that distance from the WD. Consider next the pre-shock flow: at a distance $0.1\, R_{\rm WD}$, the same ionization parameter and luminosity imply a density of around $3\times 10^{14}~\rm cm^{-3}$, and hence a volume of $0.002 \, R_{\rm WD}^3$, which is roughly the size of the accretion shock region.

Given these considerations, we constructed a model of the X-ray line emission from the photoionized flow above the stand-off shock. We assume a WD of mass $M_{\rm WD} = 0.79\, M_{\odot}$ \citep{beuerman08}, a corresponding radius $R_{\rm WD}=7.1\times 10^8$ cm (using the mass-radius relationship of \citealt{pringle}) and a luminosity $L_{\rm X}=2.6\times 10^{32}~\rm erg~s^{-1}$ at a distance of 64 pc \citep{beuerman03}. These parameters and the assumed size of the accretion spots determine the density at the shock, and this density in turn determines the cooling time and shock height.




We assume a density $n_0$ at the base of the pre-shock flow and a dipole magnetic field, hence that the pre-shock density scales with the radius $r$ as $n_0\, (R_{\rm WD}/r)^{-5/2}$. 
With this assumption, the flux of the illuminating radiation and hence the ionization parameter drops rapidly just above the shock on a length scale comparable to the shock size. The atomic physics packages described in \citet{miller} are used to compute the ionization and thermal equilibrium and the attenuation of radiation along and perpendicular to the column axis. We have explored four different models --- designated A, B, C and D --- that were chosen to match the observed luminosity, $L_{X}$, and differ only in the area of the accretion spot.  The resulting parameters of each model, listed in Table \ref{table:2}, are: the pre-shock density ($n_0$),
the optical depth for electron scattering ($\tau_{\rm e}$)
,  the size of the accretion spot ($r_{\rm spot}$), the shock height ($h_{\rm shock}$), the height of the \ion{O}{8} peak emissivity ($d$) and the observed flux of the observed broad lines. The corresponding radiative recombination continua (RRCs) should be comparable in strength to the predicted line fluxes, but they will be significantly smeared in wavelength and therefore difficult to detect. The models are quite simple in that they include only one-dimensional radiative transfer and neglect time-dependent ionization, compressional heating of the infalling gas due to the assumed dipole geometry, and density stratification in the pre-shock flow. 
However, our model should give 
the correct intensities for the strong lines.  
We note that the detection of a broad component in \ion{Fe}{17} $\lambda 16.78$, but not in $\lambda 15.01$ 
supports a photoionization origin. \citet{liedahl} predict \ion{Fe}{17} $\lambda 16.78$/\ion{Fe}{17} $\lambda 15.01 >$ 25 at densities of $10^{11}~\rm cm^{-3}$, with the ratio of both lines increasing with density. 

It should be noted, however, that the tabulated (pre-shock) densities are comparable to the post-shock densities derived by \citet{mauche01,mauche03} from the \ion{Fe}{17} and \ion{Fe}{22} emission-line ratios, whereas the post-shock density should be higher by a factor of four due to shock compression and by an additional factor due to compression as the shocked gas cools. The latter would be an order of magnitude for a steady flow, but could be much smaller due to the thermal instability of radiative shocks \citep{chevalier82}.

\begin{deluxetable*}{lccccc}
\tablecolumns{6}
\tablewidth{0pc}
\tablecaption{Photoionization model results.}
\tablehead{
\colhead{Parameter} & \colhead{Model A} & \colhead{Model B}&\colhead{Model C} & \colhead{Model D} & \colhead{Observed Flux\tablenotemark{a}}
}
\startdata
\hbox to 1.0in{$n_0$\tablenotemark{b}\leaders\hbox to 0.5em{\hss.\hss}\hfill}& 0.41&  2.7  &   12.& 30.& $\cdots$  \\
\hbox to 1.0in{$\tau_{\rm e}$\leaders\hbox to 0.5em{\hss.\hss}\hfill}& 0.03&  0.19 &  0.73& 1.8& $\cdots$  \\
\hbox to 1.0in{$r_{\rm spot}$\tablenotemark{c}\leaders\hbox to 0.5em{\hss.\hss}\hfill}& 0.098& 0.056& 0.031& 0.018& $\cdots$\\
\hbox to 1.0in{$h_{\rm shock}$\tablenotemark{d}\leaders\hbox to 0.5em{\hss.\hss}\hfill}& 0.62&  0.19 & 0.062& 0.019& $\cdots$\\
\hbox to 1.0in{$d$\tablenotemark{e}\leaders\hbox to 0.5em{\hss.\hss}\hfill}& 1.03&  0.24 & 0.080& 0.025& $\cdots$\\
\hbox to 1.0in{\oviii\tablenotemark{f}\leaders\hbox to 0.5em{\hss.\hss}\hfill}& 2.34& 2.09& 2.45& 3.08& 2.77$_{-0.3}^{+0.3}$ \\
\hbox to 1.0in{\fexviia\tablenotemark{f}\leaders\hbox to 0.5em{\hss.\hss}\hfill}& 0.21& 0.16& 0.17& 0.16& 0.55$_{-0.21}^{+0.19}$\\
\hbox to 1.0in{\mgxii\tablenotemark{f}\leaders\hbox to 0.5em{\hss.\hss}\hfill}& 0.07& 0.08& 0.10& 0.09& 0.53$_{-0.09}^{+0.10}$\\
\hbox to 1.0in{\sixiv\tablenotemark{f}\leaders\hbox to 0.5em{\hss.\hss}\hfill}& 0.03& 0.05& 0.08& 0.09& 0.84$_{-0.05}^{+0.03}$\\
\enddata
\tablenotetext{a}{For ease of comparison with the model predictions, this column reproduces the broad component flux listed in Table \ref{table:1}.}
\tablenotetext{b}{Density at the base of the pre-shock flow in units of $10^{14}~\rm cm^{-3}$.}
\tablenotetext{c}{Radius of the accretion spot on the surface of the WD in units of the white dwarf radius, $R_{\rm WD}=7.1\times 10^8$ cm.}
\tablenotetext{d}{Height of the shock above the WD surface in units of the white dwarf radius.}
\tablenotetext{e}{Height of the peak \ion{O}{8} emissivity above the WD surface in units of the white dwarf radius.} 
\tablenotetext{f}{Model predicted flux for the broad component in units of $10^{-4}$ photons cm$^{-2}$ s$^{-1}$.} 
\label{table:2}
\end{deluxetable*}

\section{Discussion and conclusions}
\label{sec:discussion}

Our photoionization models (A to D) yield similar fluxes for \ion{O}{8}, in good agreement with the observed value, but the predicted \ion{Fe}{17}, \ion{Mg}{12}, and \ion{Si}{14} fluxes are smaller by a factor of 2 to 30. Although our simple model is not accurate in the prediction of all the observed fluxes, it is appropriate in the sense that it shows that very small spots require higher pre-shock densities than those observed by \citet{mauche01,mauche03} for the post-shock flow, and very large accretion spots give small ionization parameters that cannot produce the observed flux.  Furthermore, comparison of the free-fall velocity with the line widths indicates that the accretion column must not be too far from the limb of the WD as seen from Earth. The simple model predicts that the Doppler width ($v \approx 2000~\rm km~s^{-1}$) of the ionized material is smaller than the $\sim 6000~\rm km~s^{-1}$ free-fall velocity (Section ~\ref{sec:intro}). If the columns were observed face-on, this larger Doppler shift would be seen. The smaller observed line width suggests that the accreting poles remain in the vicinity of the WD limb as seen from Earth, as supported by the small spin-phase radial velocity variations observed by \citet{ronnie2} and the small equivalent width ($\rm EW\sim 36$ eV) of the Fe K$\alpha$ line: \citet{george} calculated that such an EW would be observed if the angle between the line of sight and the reflecting surface is roughly $80^{\circ}$, i.e. the accreting poles are close to the limb.

We can decide which models are more adequate using additional arguments. First, the height of the shock should be larger than $\approx 0.015\, R_{\rm WD}$ to account for the observed modulation in the X-ray light curves \citep{allan98,ronnie1}, but less than $\sim 2\, R_{\rm WD}$ to account for the $\sim 20$ keV thermal bremsstrahlung temperature \citep{fujimoto,masses}. Second, the predicted pre-shock densities should be less than the values derived by \citet{mauche01,mauche03} from the \ion{Fe}{17} and \ion{Fe}{22} emission-line ratios. Third, a significant optical depth for electron scattering implies modulations (pulse fraction $\gtrsim$ 10\%) in the light curves in the 6--8 keV range, which were observed in {\it Ginga} data analyzed by \citet{rosen91}. \citet{allan98}, however, found only an upper limit of 4\% for the pulse fraction in the light curve in the 6--8 keV range observed with $ASCA$. Moreover, as detailed by \citet{rosen}, the favourable lines of sight to detect a high energy modulation due to electron scattering would also imply in significant photoelectric absorption in the pre-shock cold material, which will extinguish the low-energy flux and manifest itself as flat-bottomed soft X-rays spin light curves.

Based on the above conditions, we discard models C and D because: (i) their densities are higher than those obtained from the \ion{Fe}{17} and \ion{Fe}{22} \citep{mauche01,mauche03} line ratios; (ii) the combination of high densities and small accretion spot size implies significant optical depth for electron scattering (see Table \ref{table:2}), predicting flat-bottomed soft X-rays spin light curves, which are not observed \citep{ronnie1,allan98};
(iii) model D implies a small shock height, such that the collisionally ionized lines in the post-shock flow will form even closer to the white dwarf. Such emission would not show the observed sinusoidal modulation in the light curves \citep{allan98}, but would show the presence of two spikes \citep[observed in the \ion{Mg}{11} light curve, but not in the more highly ionized \ion{Mg}{12};][]{ronnie1}.

Models A and B are both acceptable in terms of: (i) their densities agree with values obtained from observations \citep{mauche01,mauche03};  
(ii) 
the optical depth for electron scattering in Models A and B is low and its contribution to the light curve modulation in the 6--8 keV energy range and absorption is small, in agreement with the absence of $\sim$100\% pulse fraction in the low-energy spin light curves and upper limit of 4\% in the pulse fraction of the light curves in the 6--8 keV range as shown by \citet{allan98}.

Summarizing, we detect for the first time in X-rays a broad component in the emission line profiles of EX~Hya and suggest that its origin is related to photoionization of a region in the pre-shock flow by radiation emitted in the collisionally ionized post-shock cooling flow. Such an origin conforms with that of the broad optical, UV, and FUV emission lines observed in EX Hya \citep{hellier87,greeley,mauche99,belle03}:
lines whose widths, fluxes, and radial velocities vary predominately on the white dwarf spin phase, which are interpreted as arising in the pre-shock flow. Even more satisfying, in two cases --- \ion{O}{6} $\lambda\lambda 1032$, 1038 in the FUV \citep{mauche99} and  \ion{N}{5} $\lambda\lambda 1239$, 1243 in the UV \citep{belle03} --- narrow components {\it that rotate with the white dwarf\/} are observed in the predominately broad emission lines. In fact, the most obvious difference between the \ion{O}{6} and \ion{N}{5} emission lines in the FUV and UV wavebands and the \oviii , \fexviia , \mgxii , and \sixiv\ emission lines in the X-ray waveband is the relative strength of the narrow and broad components; in the former case, the narrow component is the tail on the dog, whereas in the latter case it is the broad component that is the tail.
Although rather simple, our photoionization model for the origin of the broad component of the X-ray emission lines
yields appropriate values for the height of the shock and the size of the accretion spot. 
Altogether, our observational findings and the results from our model agree with a scenario where the geometry imposed by the specific accretion rate (Section ~\ref{sec:intro}) allows most of the photons from collisional excitation to escape freely, and a small fraction of those photons to photoionize the pre-shock gas. In this regard, the X-ray spectrum from EX Hya manifests both features of the ``binomial'' distribution found by \citet{koji}.


\acknowledgments
We thank the anonymous referee for useful comments which help to improve the final manuscript.
Support for this work was provided by NASA through {\it Chandra\/} awards G07-8026X issued by the {\it Chandra\/} X-ray Observatory Center, which is operated by the SAO for and on behalf of NASA under contract NAS8-03060. CWM's contribution to this work was performed under the auspices of the U.S.\ Department of Energy by Lawrence Livermore National Laboratory under Contract DE-AC52-07NA27344. NSB is supported by NASA Contract NAS8-03060 to SAO for the $Chandra$ X-ray Center.



{\it Facilities:} \facility{Chandra}

\clearpage




\begin{thebibliography}{}
\bibitem[Aizu(1973)]{aizu} Aizu, K. 1973, Prog. Theor. Phys., 49, 1184
\bibitem[Allan et al.(1998)]{allan98} Allan A., Hellier C., \& Beardmore A. P. 1998, \mnras , 295, 167
\bibitem[Belle et al.(2003)]{belle03} Belle, K. E., Howell, S. B., Sion, E. M., Long, K. S., \& Szkody, P. 2003, \apj, 587, 373
\bibitem[Beuermann et al.(2003)]{beuerman03} Beuermann, K., Harrison, Th. E., McArthur, B. E., Benedict, G. F., \& G\"ansicke, B. T. 2003, A\&A, 412, 821
\bibitem[Beuermann \& Reinsch(2008)]{beuerman08} Beuermann K., \& Reinsch K. 2008, A\&A, 480, 199
\bibitem[Brickhouse et al.(2009)]{nancy} Brickhouse, N. S. et al., 2009, ApJ, submitted.
\bibitem[Brunschweiger et al.(2009)]{masses} Brunschweiger, J., Greiner, J., Ajello, M., \& Osborne, J. 2009, A\&A, 496, 121
\bibitem[Canalle et al.(2005)]{canalle} Canalle, J. B. G., Saxton, C. J., Wu, K., Cropper, M., Ramsay, G. A., 2005, A\&A, 440, 185 
\bibitem[Chevalier \& Imamura(1982)]{chevalier82} Chevalier, R. A., \& Imamura, J. N. 1982, \apj, 261, 543 
\bibitem[Fujimoto \& Ishida(1997)]{fujimoto} Fujimoto, R. \& Ishida, M., 1997, \apj, 474, 774
\bibitem[George \& Fabian(1991)]{george} George, I. M. \& Fabian, A. C., 1991, \mnras, 249, 352
\bibitem[Greeley et al.(1997)]{greeley} Greeley, B. W., Blair, W. P., Long, K. S., Knigge, C., 1997, \apj, 488, 419
\bibitem[Hellier et al.(1987)]{hellier87} Hellier, C., Mason, K. O., Rosen, S. R., Cordova, F. A., 1987, \mnras, 228, 463
\bibitem[Hoogerwerf et al.(2005)]{ronnie2} Hoogerwerf, R., Brickhouse, N. S., \& Mauche, C. W. 2005, \apj, 628, 945
\bibitem[Hoogerwerf et al.(2006)]{ronnie1} Hoogerwerf, R., Brickhouse, N. S., \& Mauche, C. W. 2006, \apjl, 643, L45
\bibitem[Kallman \& McCray(1982)]{kallman} Kallman, T. R., \& McCray, R. 1982, \apjs , 50, 263
\bibitem[Liedahl et al.(1990)]{liedahl} Liedahl, D. A., Kahn, S. M., Osterheld, A. L., \& Goldstein, W. H. 1990, \apj, 350, 37
\bibitem[Mauche(1999)]{mauche99} Mauche, C. W. 1999, \apj , 520, 822
\bibitem[Mauche et al.(2001)]{mauche01} Mauche, C. W., Liedahl, D. A., \& Fournier, K.~B. 2001, \apj, 560, 992 
\bibitem[Mauche et al.(2003)]{mauche03} Mauche, C. W., Liedahl, D. A., \& Fournier, K.~B. 2003, \apjl, 588, L101 
\bibitem[Miller et al.(2008)]{miller} Miller, J. M., Raymond, J., Reynolds, C. S., Fabian, A. C., Kallman, T. R., \& Homan, J. 2008, \apj, 680, 1359
\bibitem[Mukai et al.(2003)]{koji} Mukai, K., Kinkhabwala, A., Paterson, J. R., Kahn S. M., \& Paerels F. 2003, \apj, 586, L77
\bibitem[Pringle \& Webbink(1975)]{pringle} Pringle, J. E. \& Webbink, R. F., 1975, MNRAS, 172, 493
\bibitem[Protassov et al.(2002)]{protassov} Protassov, R., Van Dyk, D., Connors, A., Kashyap, V. \& Siemiginowska, A., 2002, \apj, 571, 545
\bibitem[Reeves et al.(2003)]{reeves} Reeves, J. N., Watson, D., Osborne, J. P., Pounds, K. A. \& O\'Brien, P. T., 2003, A\&A, 403, 463
\bibitem[Rosen(1992)]{rosen} Rosen, S. R. 1992, \mnras , 254, 493
\bibitem[Rosen et al.(1988)]{rosen88} Rosen, S. R., Mason, K. O., \& C\'ordova, F. A. 1988, \mnras , 231, 549
\bibitem[Rosen et al.(1991)]{rosen91} Rosen, S. R., Mason, K. O., Mukai, K., \& Williams, O. R. 1991, \mnras , 249, 417
\end{thebibliography}
\end{document}